\documentclass[aps,prc,reprint,longbibliography]{revtex4-2}
\usepackage{amssymb,amsmath}
\usepackage{graphicx}
\usepackage{hyperref}
\hypersetup{
	colorlinks = true,
	linkcolor = blue,
	citecolor = blue,
	urlcolor  = blue,
	breaklinks = true,
}
\usepackage{orcidlink}

\newcommand{\erf}{\mathrm{erf}}
\newcommand{\erfc}{\mathrm{erfc}}

\renewcommand{\figurename}{Fig.}

\begin{document} 
\title{Dissipative properties of a Fermi system within the diffusion approximation of kinetic theory}
\author{Sergiy V. Lukyanov\,\orcidlink{0009-0003-0133-4483}}
\email{sergiy.lukyanov@kinr.kyiv.ua}
\affiliation{\it{Institute for Nuclear Research, 03680 Kyiv, Ukraine}}

\begin{abstract}
The dissipative properties of a Fermi system are studied within the diffusion approximation of kinetic theory for a model of a spherical atomic nucleus. 
An analytical solution of the nonlinear diffusion equation in energy space with constant kinetic coefficients is used to show that the distribution function 
asymptotically approaches the equilibrium Fermi distribution. It is found that the deviation from equilibrium at finite times decays with an effective relaxation 
time of $\tau_\text{eff}\approx 1.0\times10^{-23}$ s, whereas the asymptotic regime is characterized by an exponential decay with a relaxation time of 
$\tau_\text{eq}\approx 3\,\tau_\text{eff}$. These results explain the difference between the relaxation times extracted from integral characteristics of 
the relaxation process and from the asymptotic long-time evolution.
\end{abstract}

\maketitle
\newpage

\section{Introduction}

Dissipative phenomena in atomic nuclei play a fundamental role in the dynamics of many-body Fermi systems. They govern the evolution of nonequilibrium states, 
the transport properties of nuclear matter, and the establishment of thermodynamic equilibrium. Dissipation manifests itself in a wide range of physical processes, 
from the thermalization of excitations in heavy-ion collisions to the damping of collective modes and the description of dense nuclear matter in astrophysical environments.

One of the approaches to describing such phenomena is provided by quantum kinetic theory \cite{KoSh.b.20}. Within this framework, the evolution of nonequilibrium states 
is governed by the Landau--Vlasov kinetic equation with a collision integral on its right-hand side. Since the collision integral is analytically intractable in its 
general form, approximate methods are required for its treatment. One such method is the diffusion approximation, in which collisions are described by 
diffusion and drift processes in momentum space \cite{LiPi.bp2.81}.

In Refs.~\cite{KoLu.UJP.14,KoLu.IJMPE.15}, it was shown that, for low-temperature Fermi systems and within the small momentum-transfer approximation for particle scattering 
near the Fermi surface, the Landau--Vlasov kinetic equation can be reduced to a nonlinear diffusion equation for the Wigner distribution function in momentum space. 
Without further simplifying assumptions, this equation can only be solved numerically \cite{Ri.b.89}. To overcome this limitation, Ref.~\cite{Lu.APPB.25} introduced 
a phenomenological measure of the effective relaxation time $\tau_\text{eff}$ associated with the relaxation of an arbitrary initial excitation in momentum space 
toward equilibrium. The relaxation time is defined as the area under an equivalent exponential time dependence of the normalized integral mean-square deviation 
of the distribution function. Within the constant kinetic-coefficient approximation, this approach yields $\tau^{(p)}_{\text{eff}} \approx 8.9 \times10^{-24}$~s 
for an initially step-like distribution. An advantage of this definition is that the relaxation time is primarily determined by the early stage of the evolution, 
when the distribution function undergoes its largest changes. At the same time, it remains sensitive to the initial configuration of the system, making it possible 
to investigate the dependence of $\tau_\text{eff}$ on system parameters, such as the atomic number and the excitation energy.

In Ref.~\cite{Lu.arXiv.26}, the diffusion equation in momentum space was transformed into the corresponding equation in energy space. Since the former is 
three-dimensional whereas the latter is one-dimensional, special care must be taken to account for the density of states and to preserve particle-number conservation 
in phase space. It was shown that, within the approximation of a constant single-particle density of states, $g(\epsilon)\approx g(\epsilon_\text{F})$, the diffusion equation 
in energy space retains the same structure as the original equation in momentum space. As demonstrated in Refs.~\cite{Wo.PRL.82,BaWo.AP.19}, this nonlinear equation 
admits an exact analytical solution, whose long-time behavior is characterized by an exponential decay with a relaxation time of $\tau_\text{eq} \approx 3.2 \times10^{-23}$~s.
Although this analytical approach provides a direct determination of the relaxation time, the obtained result is asymptotic and is valid only in the long-time limit. 
Consequently, it cannot be directly applied to describe the early stages of the relaxation process.

Although the kinetic coefficients obtained in momentum space coincide, after the appropriate transformation, with those derived in energy space \cite{Lu.NPAE.23}, 
the corresponding relaxation times, $\tau_\text{eff}$ and $\tau_\text{eq}$, differ by approximately a factor of 3. A possible explanation for the origin of 
this discrepancy was suggested in Ref.~\cite{Lu.PRC.26}; however, a detailed analysis has not yet been carried out. 
This work aims to clarify the origin of the difference between the relaxation times $\tau_\text{eff}$ and $\tau_\text{eq}$.

Further analysis is carried out for a Fermi system modeling a spherical atomic nucleus in its ground state. The system is described within the infinite nuclear matter 
approximation, neglecting surface effects, so that the distribution function is independent of spatial coordinates. 

The paper is organized as follows. In Sec.~\ref{sec:exact-solution}, the diffusion equation in energy space is introduced. Within the constant kinetic-coefficient approximation, 
a solution method and the corresponding exact analytical solution are briefly outlined. Section~\ref{sec:asymp-relax} is devoted to the analysis of the long-time asymptotic limit 
of the exact solution and the temporal behavior of the deviation from equilibrium. An expression for the equilibrium relaxation time $\tau_\text{eq}$ is derived. 
In Sec.~\ref{sec:tau_eff}, the effective relaxation time $\tau_\text{eff}$ is evaluated analytically, and its behavior in the limit $t\to\infty$ is examined. 
The main conclusions are summarized in the final section.

\section{\label{sec:exact-solution} Solution of the Diffusion Equation}

Consider the nonlinear Fokker--Planck equation describing the diffusion of the distribution function in energy space,
\begin{equation}
\frac{\partial f}{\partial t} =
- \frac{\partial}{\partial \epsilon}
\left[
v f\left(1-f\right) + f^2 \frac{\partial D}{\partial \epsilon}
\right]
+ \frac{\partial^2 (f D)}{\partial \epsilon^2},
\label{eq:nonlin-FP}
\end{equation}
where $D$ and $v$ denote the diffusion and drift coefficients, respectively, which are, in general, energy dependent. In the general case, Eq.~\eqref{eq:nonlin-FP} 
can only be solved numerically \citep{Ri.b.89}. Under certain simplifying assumptions, however, it admits an exact analytical solution \cite{Wo.PRL.82,BaWo.AP.19}.

It is assumed that the kinetic coefficients are energy independent and remain constant, $D=\mathrm{const}$ and $v=\mathrm{const}$. 
Under this assumption, they can be taken outside the differential operators, and Eq.~\eqref{eq:nonlin-FP} reduces to
\begin{equation}
\frac{\partial f}{\partial t}=
-v\frac{\partial}{\partial\epsilon}\left[f(1-f)\right]
+D\frac{\partial^2 f}{\partial\epsilon^2}.
\label{eq:dif_const}
\end{equation}
The following analysis is restricted to the limit of constant kinetic coefficients.

The analytical solution of this nonlinear diffusion equation is well known \cite{Bu.b.74,Ri.b.89,Wo.PRL.82,BaWo.AP.19}. 
For completeness, one possible derivation is briefly outlined below. After expanding the nonlinear term and introducing the new function
\begin{equation}
w(\epsilon,t)=v-2v f(\epsilon,t),
\label{eq:wdef_app}
\end{equation}
Eq.~\eqref{eq:dif_const} reduces to the classical Burgers equation \cite{Bu.b.74},
\begin{equation}
\frac{\partial w}{\partial t}
+ w \frac{\partial w}{\partial \epsilon}
= D \frac{\partial^2 w}{\partial \epsilon^2}.
\label{eq:burgers_app}
\end{equation}

The Cole--Hopf transformation \cite{Ho.QAM.50,Co.b.51} is then introduced to linearize this equation
\begin{equation}
w(\epsilon,t)=-\frac{2D}{\phi(\epsilon,t)}\frac{\partial\phi(\epsilon,t)}{\partial\epsilon}.
\label{eq:colehopf_app}
\end{equation}
Substituting Eq.~\eqref{eq:colehopf_app} into Eq.~\eqref{eq:burgers_app} yields the heat equation,
\begin{equation}
\frac{\partial\phi}{\partial t}=D\frac{\partial^2\phi}{\partial\epsilon^2}.
\label{eq:heat_app}
\end{equation}

Let the initial condition be
\begin{equation}
f(\epsilon,0)=f_0(\epsilon).
\end{equation}
From Eqs.~\eqref{eq:wdef_app} and \eqref{eq:colehopf_app}, the initial condition for the function $\phi$ is obtained as
\begin{equation}
\begin{split}
\phi(\epsilon,0)&=\phi_0(\epsilon) \\
&=\exp\!\left[-\frac{1}{2D}\left(v\epsilon-2v\int_0^\epsilon f_0(x)\,dx\right)\right].
\end{split}
\label{eq:phi0_app}
\end{equation}

The solution of Eq.~\eqref{eq:heat_app} can be written as a convolution with a Gaussian \cite{Ev.b.10},
\begin{equation}
\phi(\epsilon,t)=\frac{1}{\sqrt{4\pi Dt}}
\int_{-\infty}^{\infty}
\exp\!\left[-\frac{(\epsilon-x)^2}{4Dt}\right]\phi_0(x)\,dx.
\label{eq:phi_solution_app}
\end{equation}
The distribution function is recovered as
\begin{equation}
f(\epsilon,t)=\frac12 \left[ 1+ \frac{2D}{v} \frac{\partial_\epsilon\phi}{\phi} \right].
\end{equation}
The corresponding exact integral representation is given by \cite{Wo.PRL.82}
\begin{equation}
f(\epsilon,t)=\frac{\displaystyle \int_{-\infty}^{\infty} f_0(x)\, h(x,\epsilon;t)\,dx }{\displaystyle \int_{-\infty}^{\infty} h(x,\epsilon;t)\,dx},
\label{eq:f-final}
\end{equation}
where the kernel function is defined as
\begin{equation}
\begin{split}
h(x,\epsilon;t)=\exp\!\bigg[& -\frac{(\epsilon-x)^2}{4Dt} \\
& -\frac{1}{2D} \left( v x - 2v\int_0^x f_0(y)\,dy \right) \bigg].
\end{split}
\label{eq:h_app}
\end{equation}

Equation~\eqref{eq:f-final} is the exact solution of the nonlinear diffusion equation~\eqref{eq:dif_const}. It shows that the distribution function is determined 
by a weighted average of the initial distribution $f_0(x)$ with the positive weight function $h(x,\epsilon;t)$. The nonlinearity of the original equation is entirely 
contained in the exponential factor involving the integral of $f_0(x)$ and therefore does not violate the constraint $0\le f \le 1$.

For the initial Heaviside step-function distribution,
\begin{equation}
f_0(\epsilon)=\theta(\epsilon_\text{F}-\epsilon),
\label{eq:f0}
\end{equation}
where $\epsilon_\text{F}$ is the Fermi energy, Eq.~\eqref{eq:f-final} can be expressed in terms of error functions \cite{Wo.PRL.82,BaWo.AP.19}.
\begin{equation}
\begin{split}
f(\epsilon,t)= & \left\{ 1+\erf\!\left( \dfrac{\epsilon_\text{F}-\epsilon-vt}{\sqrt{4Dt}} \right) \right\} \\
        \times & \bigg\{ 1+\erf\!\left( \dfrac{\epsilon_\text{F}-\epsilon-vt}{\sqrt{4Dt}} \right) \\
             + & \exp\!\left[ \dfrac{v(\epsilon_\text{F}-\epsilon)}{D} \right] \erfc\!\left( \dfrac{\epsilon_\text{F}-\epsilon+vt}{\sqrt{4Dt}} \right) \bigg\}^{-1}. 
\end{split}
\label{eq:f_final}
\end{equation}

\section{\label{sec:asymp-relax} Asymptotic Behavior of the Solution}

Using the odd symmetry of the error function, Eq.~\eqref{eq:f_final} can be rewritten as
\begin{equation}
f(\epsilon,t)=\frac{R(t)}{R(t)+\exp\!\left(\frac{\epsilon-\epsilon_\text{F}}{T_\text{eq}}\right)},
\end{equation}
where
\[
R(t)=\frac{\erfc(a+b)}{\erfc(a-b)}.
\]
Here,
\[
a=\frac{v}{2}\sqrt{\frac{t}{D}},
\qquad
b=\frac{\epsilon-\epsilon_\text{F}}{2\sqrt{Dt}},
\]
and
\begin{equation}
T_\text{eq}=-\frac{D}{v},
\label{eq:tempeq}
\end{equation}
where $T_\text{eq}$ is the equilibrium temperature \cite{Wo.PRL.82,Lu.PRC.26}.

Consider the limit $t\to\infty$ for finite $\epsilon$. For constant kinetic coefficients, the effective drift coefficient is negative, $v=-D/T_\text{eq}$. Therefore,
\begin{equation}
a=\frac{v}{\sqrt{4D}}\sqrt{t}\to -\infty ,
\qquad
b=\frac{\epsilon - \epsilon_\text{F}}{\sqrt{4Dt}}\to 0.
\label{eq:lim-ab_1}
\end{equation}

Since the derivative of the complementary error function is given by \cite{AbSt.b.72}
\[
\frac{d\, \erfc(z)}{dz}=-\frac{2}{\sqrt{\pi}}e^{-z^2},
\]
a Taylor expansion about $a$ in the small parameter $b$ yields
\[
\erfc(a\pm b)\approx \erfc(a) \mp \frac{2}{\sqrt{\pi}}e^{-a^2} b.
\]

Thus, the ratio can now be approximated as
\[
\begin{split}
R(t)&=\frac{\erfc(a+b)}{\erfc(a-b)} \approx \frac{\erfc(a) - \frac{2b}{\sqrt{\pi}} e^{-a^2}}{\erfc(a) + \frac{2b}{\sqrt{\pi}} e^{-a^2}} \\
& \approx 1-\frac{\frac{4b}{\sqrt{\pi}}e^{-a^2}}{\erfc(a) + \frac{2b}{\sqrt{\pi}} e^{-a^2}} \approx 1 - \frac{4b}{\sqrt{\pi}\; \erfc(a)} e^{-a^2} \\
& = 1 - r(t).
\end{split}
\]

For large negative arguments, the standard asymptotic expansion of the complementary error function is used \cite{AbSt.b.72}:
\begin{equation}
\erfc(z)\approx 2-\frac{e^{-z^2}}{\sqrt{\pi}|z|}.
\end{equation}
It then follows that
\[
r(t) = \frac{4}{\sqrt{\pi}\, \erfc(a)}\, b\, e^{-a^2}
\approx \frac{2}{\sqrt{\pi}}\, b\, e^{-a^2}.
\]
Substituting the expressions for $a$ and $b$ gives
\begin{equation}
\label{eq:r-def}
r(t) \approx \frac{\epsilon-\epsilon_\text{F}}{\sqrt{\pi D}}\, t^{-1/2} \exp\!\left(-\frac{t}{\tau_\text{eq}}\right),
\end{equation}
where
\begin{equation}
\tau_\text{eq}=\frac{4D}{v^2}
\label{eq:tau_eq}
\end{equation}
is the equilibrium relaxation time \cite{Wo.PRL.82,Ev.b.10}.

According to Eq.~\eqref{eq:r-def}, $r$ vanishes in the limit $t\to\infty$, and the distribution function approaches the equilibrium Fermi distribution
\begin{equation}
f_\text{eq}(\epsilon)=\left[1+\exp\left(\frac{\epsilon-\epsilon_\text{F}}{T_\text{eq}}\right)\right]^{-1}.
\label{eq:f_Fermi}
\end{equation}

The deviation of the distribution function from its equilibrium value is introduced as
\begin{equation}
\delta f(\epsilon,t) = f(\epsilon,t) - f_\text{eq}(\epsilon).
\label{eq:df-def}
\end{equation}
Using Eq.~\eqref{eq:r-def}, the following asymptotic expression for $\delta f(\epsilon,t)$ is obtained
\begin{equation}
\label{eq:df-final}
\delta f(\epsilon,t) = - f_\text{eq}(\epsilon)\, \frac{\epsilon-\epsilon_\text{F}}{\sqrt{\pi D}}\, t^{-1/2} \exp\!\left(-\frac{t}{\tau_\text{eq}}\right).
\end{equation}

\begin{figure}[!b]
\includegraphics[width=0.98\columnwidth,clip]{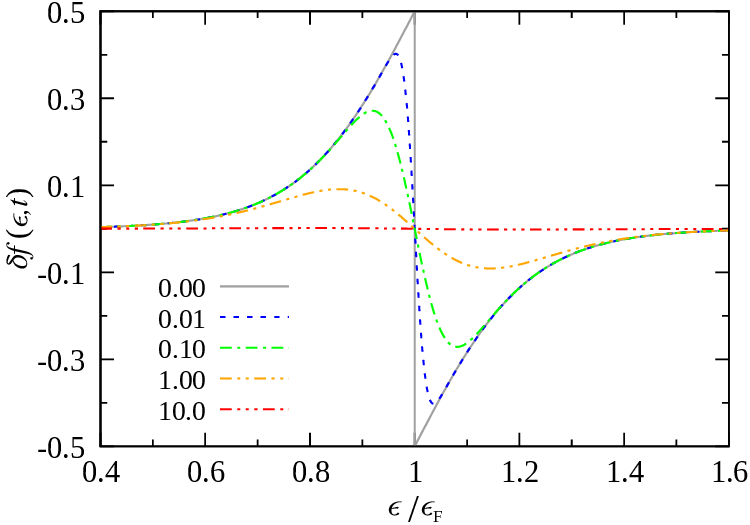}
\caption{Difference $\delta f(\epsilon,t)$ between the exact distribution function~\eqref{eq:f_final} and the equilibrium Fermi distribution~\eqref{eq:f_Fermi} 
versus the dimensionless energy $\epsilon/\epsilon_\text{F}$. The curves correspond to the dimensionless times $t/(10^{-23}\,\mathrm{s})=0.00,\ 0.01,\ 0.10,\ 1.00,\ 10.0$.}
\label{fig:1}
\end{figure}

\figurename~\ref{fig:1} shows the results of numerical calculations for the difference $\delta f(\epsilon,t)$ between the exact distribution function~\eqref{eq:f_final} 
and the equilibrium Fermi distribution~\eqref{eq:f_Fermi} as a function of the dimensionless energy $\epsilon/\epsilon_\text{F}$ at different times.

The Fermi energy was taken to be $\epsilon_\text{F}=37$ MeV. The following values are adopted for the diffusion and drift coefficients~\cite{Wo.PRL.82,BaWo.AP.19}:
\[
D=20\times 10^{23}\,\mathrm{MeV}^{2}\,\mathrm{s}^{-1},
\qquad
v=-5\times 10^{23}\,\mathrm{MeV}\,\mathrm{s}^{-1}.
\]
According to Eqs.~\eqref{eq:tempeq} and~\eqref{eq:tau_eq}, these values correspond to the equilibrium temperature $T_\text{eq}=4$ MeV and the equilibrium relaxation time 
$\tau_\text{eq}\approx3.2\times10^{-23}$ s. They therefore satisfy the conditions of the low-temperature approximation.

The solid curve shows the difference between the initial step-function distribution~\eqref{eq:f0} and the equilibrium Fermi distribution~\eqref{eq:f_Fermi}. 
As seen in the figure, the initial deviation is localized primarily in the vicinity of the Fermi surface and has opposite signs below and above $\epsilon_\text{F}$.
With increasing time, the deviation decreases monotonically and becomes practically negligible for $t > 10\times 10^{-23}$~s.

\section{\label{sec:tau_eff}Effective Relaxation Time}

A direct analysis of the asymptotic deviation in Eq. \eqref{eq:df-final} is cumbersome and not particularly informative. 
It can therefore be recast in a more transparent form. To this end, the definition of the effective relaxation time introduced 
in Ref.~\cite{Lu.APPB.25} is adopted, but in energy space:
\begin{equation}
\tau_\text{eff}= \int_0^\infty \frac{\Delta (t)}{\Delta_0} dt,
\label{eq:taueff-def}
\end{equation}
where
\begin{equation}
\Delta (t) = \sqrt{\int_0^\infty \left[ \delta f(\epsilon,t) \right]^2 d\epsilon}
\label{eq:delta-def}
\end{equation}
is the root-mean-square deviation, while
\begin{equation}
\Delta (t=0)\equiv\Delta_{0}= \sqrt{\int_0^\infty \left[ \delta f_\text{in}(\epsilon) \right]^2 d\epsilon},
\label{eq:delta0-def}
\end{equation}
gives its initial value. 
It is worth noting that, in the approximation $g(\epsilon)\approx g(\epsilon_F)$, Eqs.~\eqref{eq:taueff-def}--\eqref{eq:delta0-def} 
are equivalent to the effective relaxation time defined in momentum space in Ref.~\cite{Lu.APPB.25}.

For the initial step-function distribution~\eqref{eq:f0}, the quantity $\Delta_0$ can be evaluated analytically. 
To this end, the initial deviation of the distribution function is written as
\begin{equation}
\begin{split}
\delta f_{\text{in}}(\epsilon) & =f_0(\epsilon)-f_{\text{eq}}(\epsilon) \\
                               & =\Theta(\epsilon_\text{F}-\epsilon)-\frac{1}{1+e^{(\epsilon-\epsilon_\text{F})/T_{\text{eq}}}}.
\end{split}
\label{eq:deltafin}
\end{equation}

Substituting Eq.~\eqref{eq:deltafin} into Eq.~\eqref{eq:delta0-def} yields
\begin{equation}
\Delta_0^2= \int_0^\infty\left[\Theta(\epsilon_\text{F}-\epsilon)-\frac{1}{1+e^{(\epsilon-\epsilon_\text{F})/T_{\text{eq}}}}\right]^2d\epsilon.
\label{eq:delta0-squared}
\end{equation}

To evaluate the integral in Eq.~\eqref{eq:delta0-squared}, the dimensionless variable $x=(\epsilon-\epsilon_\text{F})/T_\text{eq}$ is introduced. 
After the change of variables, the integral takes the form
\begin{equation}
\Delta_0^2= T_{\text{eq}} \int_{-\epsilon_\text{F}/T_\text{eq}}^{\infty} \left[ \Theta(-x)-\frac{1}{1+e^x} \right]^2 dx.
\label{eq:delta0-squared-x}
\end{equation}

In the low-temperature limit, $T_\text{eq}\ll\epsilon_\text{F}$, the lower limit of integration can be replaced by $-\infty$. 
For $x<0$, the integrand in Eq.~\eqref{eq:delta0-squared-x} becomes
\begin{equation}
\Theta(-x)=1, \qquad 1-\frac{1}{1+e^x} = \frac{e^x}{1+e^x},
\label{eq:teta_xlt0}
\end{equation}
whereas for $x>0$
\begin{equation}
\Theta(-x)=0.
\label{eq:teta_xgt0}
\end{equation}

The integral can then be decomposed into the sum of two contributions
\begin{equation}
I=\int_{-\infty}^0\left(\frac{e^x}{1+e^x}\right)^2 dx+\int_0^\infty\left(\frac{1}{1+e^x}\right)^2 dx.
\label{eq:Int}
\end{equation}
Because of the symmetry of the logistic function, the two integrals are equal, and it is therefore sufficient to evaluate only one of them. 
This yields
\begin{equation}
I=2\ln(2)-1.
\label{eq:Int-approx}
\end{equation}

Substituting the result of Eq.~\eqref{eq:Int-approx} into Eq.~\eqref{eq:delta0-squared-x} gives the final expression for the squared initial root-mean-square deviation
\begin{equation}
\Delta_0^2=(\ln(4)-1)\, T_{\text{eq}}.
\label{eq:delta0-squared-final}
\end{equation}

\begin{figure}[!t]
\includegraphics[width=0.98\columnwidth,clip]{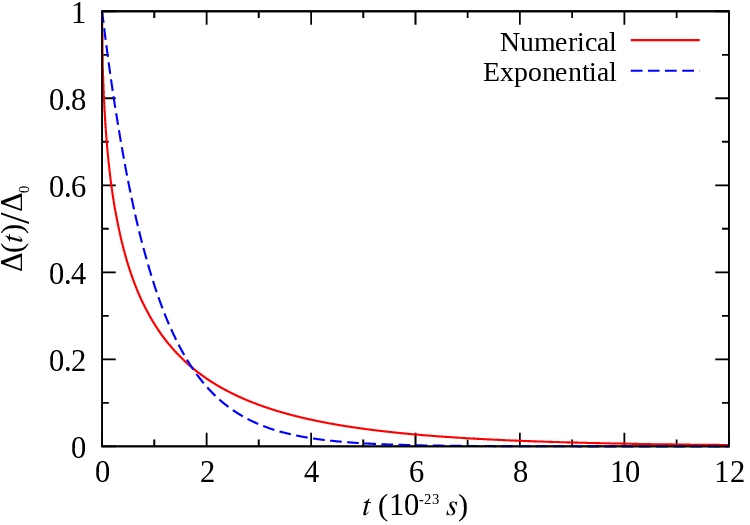}
\caption{Time dependence of the ratio $\Delta(t)/\Delta_0$. The solid curve represents the numerical result, while the dashed curve corresponds 
to an exponential decay with the relaxation time $\tau=\tau_\text{eff}\approx 1.0\times 10^{-23}$~s.}
\label{fig:2}
\end{figure}

The quantity $\Delta(t)$ is evaluated numerically. To this end, the definition of the deviation from equilibrium, Eq.~\eqref{eq:df-def}, together with the expressions 
for the equilibrium Fermi distribution, Eq.~\eqref{eq:f_Fermi}, and the exact distribution function, Eq.~\eqref{eq:f_final}, is employed. 
The resulting time dependence of the ratio $\Delta(t)/\Delta_0$ is shown by the solid curve in \figurename~\ref{fig:2}.
As seen in \figurename~\ref{fig:2}, the ratio $\Delta(t)/\Delta_0$ is equal to $1$ at $t=0$, as follows directly from its definition. 
It then decreases with time, exhibiting behavior similar to that reported in Ref.~\cite{Lu.APPB.25}.

Integration of this dependence over time according to the definition of the effective relaxation time, Eq.~\eqref{eq:taueff-def}, yields the effective relaxation time 
characterizing the dissipation process. As seen in \figurename~\ref{fig:2}, the main contribution to the integral arises from relatively short times, $t<10\times 10^{-23}$~s. 
The resulting value, $\tau_\text{eff}\approx 1.0\times 10^{-23}$~s, is in good agreement with the value $\tau^{(p)}_\text{eff}\approx 8.9\times 10^{-24}$~s obtained 
in momentum space in Ref.~\cite{Lu.APPB.25}. The dashed curve in \figurename~\ref{fig:2} 
represents the exponential decay $\exp(-t/\tau)$ with the relaxation time $\tau=\tau_\text{eff}\approx 1.0\times 10^{-23}$~s. A comparison with the exact dependence 
$\Delta(t)/\Delta_0$ shows that the relaxation is nonexponential at short times.

Since the asymptotic expression for $\delta f$, Eq.~\eqref{eq:df-final}, is known analytically and exhibits an exponential decay with the characteristic relaxation time 
$\tau_\text{eq}$, the corresponding asymptotic behavior of $\Delta(t)/\Delta_0$ can also be derived.

Substituting the asymptotic expression~\eqref{eq:df-final} into the definition~\eqref{eq:delta-def} yields the following expression 
for the squared root-mean-square deviation:
\begin{equation}
\Delta^2(t) = \frac{1}{\pi D}\; \frac{e^{-2t/\tau_\text{eq}}}{t}
\int_0^{\infty} \left[ \frac{\epsilon-\epsilon_\text{F}}{1+e^{(\epsilon-\epsilon_\text{F})/T_{\text{eq}}}} \right]^2 d\epsilon.
\label{eq:deltat-squared}
\end{equation}

Introducing the dimensionless variable $x$ as before and changing the integration variable yields
\begin{equation}
\Delta^2(t) = \frac{T^3_\text{eq}}{\pi D}\; \frac{e^{-2t/\tau_\text{eq}}}{t}
\int_{-\epsilon_\text{F}/T_\text{eq}}^{\infty} \left[ \frac{x}{1+e^{x}} \right]^2 dx.
\label{eq:deltat-squared-x}
\end{equation}

The integral in Eq.~\eqref{eq:deltat-squared-x} is evaluated in the low-temperature limit
\begin{equation}
\int_{-\epsilon_\text{F}/T_\text{eq}}^{\infty}\frac{x^2\, dx}{[1+e^x]^2} \approx \frac13\left(\frac{\epsilon_\text{F}}{T_\text{eq}}\right)^3.
\label{eq:I2}
\end{equation}
Substituting this result yields the asymptotic expression
\begin{equation}
\Delta^2(t) \approx \frac{\epsilon_\text{F}^3}{3\pi D}\; t^{-1}e^{-2t/\tau_\text{eq}} .
\label{eq:deltat-squared-x-1}
\end{equation}

Normalizing this expression by Eq.~\eqref{eq:delta0-squared-final} and taking the square root yields
\begin{equation}
\frac{\Delta(t)}{\Delta_0} \approx  \frac{\epsilon_\text{F}}{\sqrt{3(\ln(4)-1)\pi D}}
\sqrt{\frac{\epsilon_\text{F}}{T_\text{eq}}}\; t^{-1/2}e^{-t/\tau_\text{eq}}.
\label{eq:frac-deltat0-approx}
\end{equation}
The obtained expression is valid in the asymptotic limit $t\to\infty$ and in the low-temperature limit, $T_\text{eq}\ll\epsilon_\text{F}$.

\begin{figure}[!b]
\includegraphics[width=0.98\columnwidth,clip]{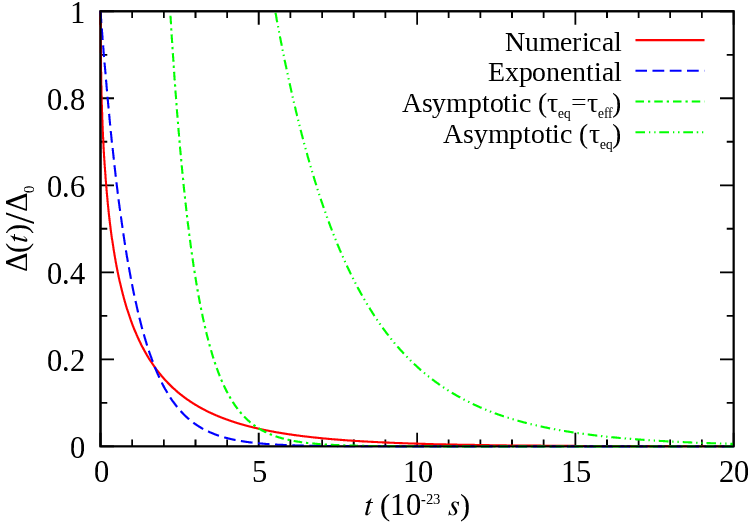}
\caption{Time dependence of the ratio $\Delta(t)/\Delta_0$. The solid and dashed curves are reproduced from \figurename~\ref{fig:2}. 
The dash-dotted curve corresponds to the asymptotic expression~\eqref{eq:frac-deltat0-approx} with the relaxation time chosen as 
$\tau_\text{eq}=\tau_\text{eff}\approx 1.0\times 10^{-23}$~s,
whereas the dash-double-dotted curve corresponds to the same asymptotic expression with the relaxation time $\tau_\text{eq}\approx 3.2\times 10^{-23}$~s.}
\label{fig:3}
\end{figure}

For comparison, \figurename~\ref{fig:3} also shows the asymptotic dependence~\eqref{eq:frac-deltat0-approx}, represented by the dash-double-dotted curve with 
the relaxation time $\tau_\text{eq}\approx 3.2\times 10^{-23}$~s, together with the results shown in \figurename~\ref{fig:2}.
As seen in \figurename~\ref{fig:3}, the asymptotic expression~\eqref{eq:frac-deltat0-approx} lies outside its range of validity, at least over the time interval 
$t<5.6\times10^{-23}$~s, since it predicts values exceeding $1$. As time increases, the asymptotic expression decays significantly more slowly than the exact result. 
As pointed out in Ref.~\cite{Lu.PRC.26}, this difference reflects the distinct physical meaning of the two time scales: $\tau_\text{eff}$ characterizes the relaxation 
of the system at finite times, whereas $\tau_\text{eq}$ determines the asymptotic behavior of the exact analytical solution of the diffusion equation in the limit 
$t\to\infty$.

To verify this conclusion, an additional calculation based on Eq.~\eqref{eq:frac-deltat0-approx} was performed, in which the relaxation time was set to 
$\tau_\text{eq}=\tau_\text{eff}\approx 1.0\times 10^{-23}$~s instead of $3.2\times10^{-23}$~s. 
The corresponding dependence is shown by the dash-dotted curve. It lies considerably closer to the exact result shown in \figurename~\ref{fig:2}, 
but still exceeds the physically admissible range for $t<2.3\times 10^{-23}$~s. 
At asymptotically large times, however, this curve lies below the dash-double-dotted one, in agreement with the conclusions of Ref.~\cite{Lu.PRC.26}.

\begin{figure}[!b]
\includegraphics[width=0.98\columnwidth,clip]{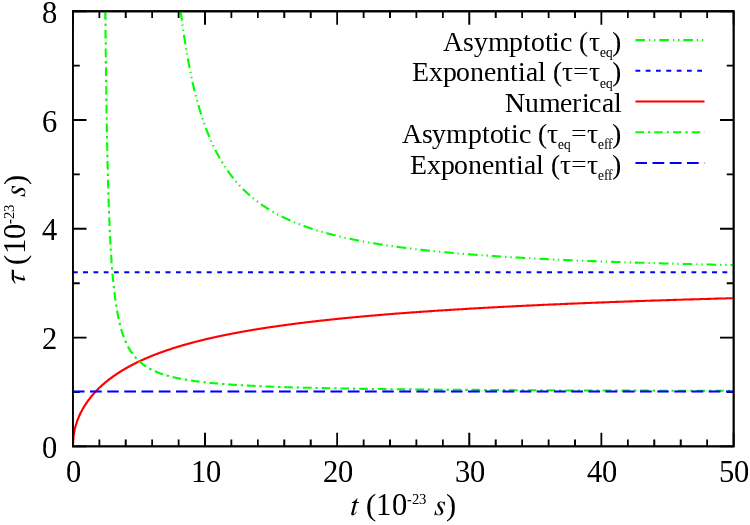}
\caption{Time dependence of the quantity $\tau$. The calculation parameters and curve styles are the same as those used in the caption of \figurename~\ref{fig:3}. 
In addition, the short-dashed horizontal line corresponds to the exponential dependence with the relaxation time $\tau=\tau_\text{eq}\approx 3.2\times 10^{-23}$~s.}
\label{fig:4}
\end{figure}

To compare the asymptotic behavior of the ratio $\Delta(t)/\Delta_0$ for the different curves shown in \figurename~\ref{fig:3}, it is convenient to eliminate 
the exponential time dependence by introducing the quantity
\begin{equation}
\tau(t)=-\frac{t}{\ln\left(\Delta(t)/\Delta_0\right)},
\label{eq:tau}
\end{equation}
which can be interpreted as an instantaneous relaxation time. In particular, for the purely exponential curve shown by the dashed line in \figurename~\ref{fig:3}, 
the quantity $\tau$ is constant and equal to $\tau=\tau_\text{eff}\approx 1.0\times 10^{-23}$~s, as follows directly from the definition of the exponential relaxation law. 
The corresponding curve is shown in \figurename~\ref{fig:4} by the horizontal dashed line.
The remaining curves were constructed in the same manner. As seen in \figurename~\ref{fig:4}, the quantity $\tau$ for the exact solution (solid curve) depends on time, 
indicating the nonexponential character of the relaxation process. As time increases, it grows monotonically and approaches the value 
$\tau_\text{eq}\approx 3.2\times 10^{-23}$~s from below. In contrast, for the asymptotic solution~\eqref{eq:frac-deltat0-approx}, the quantity $\tau$ approaches the same value 
from above. This behavior can be demonstrated analytically. Extracting the time dependence from Eq.~\eqref{eq:frac-deltat0-approx} and absorbing all remaining factors into 
the constant $B$, substitution into Eq.~\eqref{eq:tau} yields
\begin{equation*}
\tau(t)=\frac{\tau_\text{eq}}{1-\frac{\tau_\text{eq}}{t}\left(\ln(B)-\frac12\ln(t)\right)}.
\end{equation*}
Hence, in the limit $t\to\infty$,
\begin{equation*}
\tau(t)\approx\tau_\text{eq}
+\tau_\text{eq}^2\frac{\ln(B)-\frac12\ln(t)}{t}
+ \mathcal{O} \left(\frac{\ln^2(t)}{t^2}\right).
\end{equation*}
Thus, $\tau(t)$ indeed approaches $\tau_\text{eq}$, but only slowly, with a correction of order $\ln(t)/t$.

Thus, in the limit $t\to\infty$, both the exact and the asymptotic solutions converge to the same asymptotic relaxation time $\tau_\text{eq}$. At the same time, 
the asymptotic solution~\eqref{eq:frac-deltat0-approx} evaluated with the relaxation time set to $\tau_\text{eq}=\tau_\text{eff}\approx 1.0\times 10^{-23}$~s 
(dash-dotted curve) gradually approaches the exponential dependence with the same relaxation time (dashed curve) as time increases. Therefore, the quantity 
$\tau_\text{eff}$ characterizes the initial stage of the relaxation process, whereas the asymptotic regime is governed by the substantially longer relaxation time $\tau_\text{eq}$.

The above results provide an explanation for the difference between the relaxation times obtained by different methods. The quantity $\tau_\text{eff}$, defined in terms 
of the integral characteristics of the deviation of the distribution function, is sensitive to the behavior of the system at finite times and therefore characterizes 
the effective relaxation rate over the entire evolution process. In contrast, the parameter $\tau_\text{eq}$ is determined by the asymptotic behavior of the solution 
in the limit $t\to\infty$ and characterizes only the long-time exponential relaxation regime.

Thus, the quantities $\tau_\text{eff}$ and $\tau_\text{eq}$ characterize different temporal regimes of the relaxation process and therefore do not coincide 
in systems exhibiting nonexponential relaxation dynamics.

\section{Conclusions}

The dissipative properties of a Fermi system modeling a spherical atomic nucleus in its ground state have been investigated. 
The analysis was carried out within the infinite nuclear matter approximation, neglecting surface effects. 
The consideration was restricted to the low-temperature limit, $T_\text{eq}\ll\epsilon_\text{F}$.

Within the diffusion approximation of kinetic theory and under the assumption of a constant density of single-particle states, 
the time evolution of the distribution function can be conveniently described by a nonlinear diffusion equation in energy space. 
For constant kinetic coefficients, this equation is considerably simplified and admits an analytical solution. 
Two alternative derivations of this solution are available, one of which is outlined in the present work. 
For the initial step distribution, an exact analytical expression for the distribution function in terms of error functions has been obtained. 
It has been shown that, in the limit $t\to\infty$, the solution converges to the equilibrium Fermi distribution with the temperature $T_\text{eq}=4$~MeV, 
determined by the ratio of the diffusion coefficient, $D=20\times10^{23}\,\mathrm{MeV}^2\,\mathrm{s}^{-1}$, to the drift coefficient, 
$v=-5\times10^{23}\,\mathrm{MeV}\,\mathrm{s}^{-1}$.

The deviation $\delta f$ between the exact solution and the equilibrium distribution function has been analyzed. 
Numerical simulations show that this deviation gradually decreases with time and becomes practically negligible for $t>10\times 10^{-23}$~s. 
An analytical expression for $\delta f$ has been derived in the limit $t\to\infty$, demonstrating an exponential decay with the characteristic relaxation time 
$\tau_\text{eq}=4D/v^2\approx 3.2\times 10^{-23}$~s, determined by the kinetic coefficients of the system.

It has been shown that, for the initial step distribution, the quantity $\Delta_0$ can be evaluated analytically, 
with its value being proportional to the equilibrium temperature. 
The quantity $\Delta(t)$ has been calculated numerically from the obtained deviation $\delta f$. 
It has been demonstrated that the main contribution to the time integral defining the effective relaxation time $\tau_\text{eff}$ 
arises from relatively short time scales, $t<10\times 10^{-23}$~s. 
The time dependence of the ratio $\Delta(t)/\Delta_0$ has been shown to be nonexponential in the short-time regime. 
Thus, the quantity $\tau_\text{eff}$ characterizes the relaxation of the system on finite time scales and is therefore sensitive to the configuration of the initial state.

The effective relaxation time $\tau_\text{eff}$, defined in terms of the normalized integral root-mean-square deviation of the distribution function, has been evaluated. 
For the initial step distribution, its value is found to be $\tau_\text{eff}\approx 1.0\times 10^{-23}$~s. 
This result is in good agreement with the value $\tau_\text{eff}^{(p)}\approx 8.9\times 10^{-24}$~s obtained in momentum space~\cite{Lu.APPB.25}.

Based on the asymptotic expression for $\delta f$, a corresponding analytical expression for $\Delta(t)$ has been derived. 
The resulting ratio $\Delta(t)/\Delta_0$ shows that, over the time interval $t<5.6\times 10^{-23}$~s, it falls outside its range of validity by taking values 
greater than $1$. This demonstrates the limitation of the asymptotic approach, which is applicable only in the long-time limit.

To analyze the asymptotic behavior of the quantities entering the definition of the relaxation time, the quantity $\tau(t)$ has been introduced. 
It eliminates the exponential time dependence through the logarithm of the ratio $\Delta(t)/\Delta_0$, and the resulting inverse quantity 
multiplied by $-t$ can be interpreted as an instantaneous relaxation time. 
Numerical analysis confirms that this quantity is constant for purely exponential relaxation. 
The calculated dependence $\tau(t)$ exhibits a monotonic increase with time and approaches the value $\tau_\text{eq}\approx 3.2\times 10^{-23}$~s from below. 
In contrast, for the asymptotic solution~\eqref{eq:frac-deltat0-approx}, $\tau(t)$ approaches the same value from above. 
It has been shown analytically that $\tau(t)$ converges slowly to $\tau_\text{eq}$, with a correction of order $\ln(t)/t$.

Thus, in the limit $t\to\infty$, both the exact and the asymptotic solutions converge to the same asymptotic relaxation time, $\tau_\text{eq}\approx 3.2\times10^{-23}$~s. 
At the same time, the asymptotic solution evaluated with $\tau_\text{eq}= 1.0\times 10^{-23}$~s gradually approaches purely exponential behavior 
with the same characteristic relaxation time as $t$ increases. 
Therefore, the quantity $\tau_\text{eff}$ describes the initial stage of the relaxation process, 
whereas the asymptotic regime is governed by the longer characteristic time $\tau_\text{eq}$.

The above results explain the difference between the relaxation times obtained by different methods. 
The quantity $\tau_\text{eff}$, defined in terms of the integral characteristics of the deviation of the distribution function, 
is sensitive to the behavior of the system on finite time scales and characterizes the effective relaxation rate throughout the entire evolution process. 
In contrast, the parameter $\tau_\text{eq}$ is determined by the asymptotic behavior of the solution in the limit $t\to\infty$ 
and describes only the long-time exponential relaxation regime. 
Thus, the quantities $\tau_\text{eff}$ and $\tau_\text{eq}$ characterize different temporal regimes of relaxation 
and therefore do not generally coincide in systems exhibiting nonexponential relaxation dynamics.

\bibliography{references}

\end{document}